\documentclass[preprint, amsmath,amssymb, prl]{revtex4-1}
\usepackage{graphicx}% Include figure files
\usepackage{dcolumn}% Align table columns on decimal point
\usepackage{bm}% bold math
\usepackage{hyperref}% add hypertext capabilities
\usepackage{xcolor}
\hypersetup{
    colorlinks=true,
    linkcolor=blue,
    filecolor=magenta,
    citecolor = blue,
    urlcolor=blue,
}
\usepackage{caption}
\usepackage{subcaption}
\graphicspath{{Pictures/}}
\begin{document}

%\preprint{APS/123-QED}

\title{Sub-nanometer optical linewidth of thulium atoms in rare gas crystals}

\author{Vinod Gaire}
 %\email{vgaire3@gatech.edu}
 \author{Chandra S. Raman}
 %\email{chandra.raman@physics.gatech.edu}
\author{Colin V. Parker}%
 \email{cparker@gatech.edu}
\affiliation{%
 School of Physics, Georgia Institute of Technology, Atlanta, Georgia, 30332, USA\\
 }

%\date{July, 2018}
\date{\today}
%\pacs{}% PACS, the Physics and Astronomy

\begin{abstract}
We investigate the 1140 nm magnetic dipole transition of thulium atoms trapped in solid argon and neon. These solids can be straightforwardly grown on any substrate at cryogenic temperatures, making them prime targets for surface sensing applications. Our data are well described by a splitting of the single vacuum transition into three components in both argon and neon, with each component narrower than the 0.8 nm spectrometer resolution. The lifetime of the excited states is 14.6(0.5) ms in argon and 27(3) ms in neon, shorter than in vacuum or in solid helium. We also collected visible laser-induced fluorescence spectroscopy showing broader emission features in the range of 580-600 nm. The narrow infrared features in particular suggest a range of possible applications.
\end{abstract}
\maketitle


%\tableofcontents

\section{Introduction}

Isolated atoms are narrow-linewidth optical emitters that make very precise clocks \cite{Helmcke2003, Ushijima2015, RevModPhys.87.637} and sensors \cite{Budker2007}.
%Critical to realizing the full potential of atomic devices is the manipulation of the quantum state of the atom using light, for example, by optical pumping into a known initial state or using optical pulses to create quantum superpositions.
However, realizing this isolation requires that the density be low and the atoms be far from surfaces.
%This limitation stems both from the length scale of atomic confinement schemes (typically 100's of nanometers in the best case), and from their strength (much weaker than van der Waals interactions at the nanometer scale).
On the other hand, atoms or atom-like defects in solid systems retain many of the magnetic and optical properties of free atoms, including in some cases long spin coherence times and narrow homogeneous linewidths\cite{Ishikawa1997,Riedmatten2008,Robledo2011,Raston:2013}. Unlike isolated atoms, these systems are confined in their hosts, which affords a wealth of opportunities for quantum sensing and quantum control at very short length scales (1-10 nm) proximate to surfaces. Many interesting phenomena in exotic materials occur at these scales. For example, breaking of spatial symmetry occurs at the nanoscale in cuprates\cite{Parker:2010,Comin:2014}, pnictides\cite{Rosenthal:2014}, and dichalcogenides\cite{Arguello:2015}, and the interplay with nanoscale inhomogeneity is important. Spin liquids also feature a wealth of interesting physics at this length scale\cite{Savary:2017}. 

The best techniques to investigate these length scales currently are scattering probes and scanning tunneling microscopy/spectroscopy (STM/STS). Few- or even single-atom probes of these systems would be a useful complement, since atoms are influenced by the nearby magnetic, electric, and chemical environment rather than the electronic density of states in the material. Furthermore, each atom is identical, so that one does not have to determine if the measured properties are characteristic of the material or the specifics of the probe tip\cite{Neto:2013}. Even quantum entanglement can be probed atomically, by allowing the sensor atoms to entangle with the system of interest or with each other. Solid state optical emitters can also be used for photon storage, to couple quantum information between light and matter\cite{Riedmatten2008}. However, all these possibilities rely on controlling the atomic state with good fidelity, something that is hindered if broadening prevents the resolution of internal spin, hyperfine, and crystal field states of the trapped atoms. A limited set of solid state optical emitters are narrow enough to be used, with one very successful example being nitrogen-vacancy (NV) centers in diamond\cite{Rondin2014}, where high fidelity readout and quantum operations are possible\cite{Robledo2011}.

We note that the range of narrow linewidth solid state optical emitters may actually be quite large, if one considers solids formed from noble gases. Embedding optically active species in inert solids, also known as matrix isolation, is frequently used in chemistry to study unstable or reactive species \cite{Dunkin1998,matrixIsolation}. Noble gas solids are chemically inert and easy to produce, therefore they have potential applications where the substrate must avoid surface charging. Although alkali atomic lines are typically broadened to around 100 nm when trapped in argon \cite{fajardo1993,kunttu1999,Kanagin2013}, matrix isolated molecular transitions can be narrow, and mid-infrared spin-orbit lines in atomic bromine have less than $1 \textrm{ cm}^{-1}$ linewidth\cite{Raston:2013}.

Motivated by these applications we investigate near-infrared inner shell $f$-$f$ transitions of thulium in argon and neon. Previous efforts have shown that the magnetic dipole line near 1140 nm in thulium remains narrow (less than 0.2 nm) when it is trapped in solid helium \cite{Ishikawa1997}. However, for many applications solid helium is undesireable as it can only form under high pressure. Argon and neon on the other hand can be easily grown on any substrate. We find that the line is split into at least two and likely three components, but otherwise unshifted from the vacuum wavelength of 1140 nm, and remains narrower than the spectrometer resolution of 0.8 nm. The natural linewidth in vacuum is 1.6 Hz\cite{Golovizin2015}, and it remains an open question what the true linewidth is in any noble gas. We have also determined the decay lifetimes, which are 28(3) ms and 14.6(5) ms when trapped in neon and argon, respectively, with the parentheses indicating statistical uncertainty. These are to be compared with a 75(3) ms lifetime in helium \cite{Ishikawa1997}.

The spectra of lanthanides are interesting because of the presence of unfilled $4f$ shell electrons (except in the case of neutral Yb). The ground state $4f^n6s^2$ of the neutrals is split into fine-structure ``sub levels'' with transitions at optical frequencies. Because the incomplete $4f$ shell is submerged inside the $5s$, $5p$, and $6s$ shells, the $f$-$f$ transitions are relatively isolated from the trapping environment. It is important to understand the mechanism by which the solid state confinement causes broadening. Basically, the trapping environment for the optically excited state is distinct from that of the ground state. Hence, a decay from the optically excited state proceeds by photon emission and subsequent relaxation of the environment by lattice phonons. Therefore, just as for two photon emission it is not the case that the optical linewidth is connected to the lifetime. Indeed, lifetimes for neutral Yb trapped in neon can be hundreds of seconds \cite{Welp2011}.

Thulium represents a simple case of the unfilled $f$ shell. There is only one natural isotope (${}^{169}\textrm{Tm}$ with $I = 1/2$), and the ground  electronic configuration $[\textrm{Xe}]4f^{13}6s^{2}$ has only two fine structure levels, the ${}^2F_{7/2}$ ground state and ${}^2F_{5/2}$. These levels are coupled by a magnetic dipole transition at 1140 nm. Early experiments on thulium trapped in solid neon at 4 K \cite{Belyaeva1986} showed absorption and emission bands, which were attributed to electric dipole transitions associated with $4f$-$5d$ and $6s$-$6p$ in the visible range \cite{Martin}. It has been shown that the magnetic dipole transition at 1140 nm is shielded significantly by the outer shells in high pressure helium gas \cite{Aleksandrov1984} and solid helium \cite{Ishikawa1997}. Even though the collisional transfer of angular momentum in Tm-He collisions was very small, the spin relaxation rates for Tm-Tm collisions was found to be large \cite{Connolly2010}, ruling out the possibility of evaporative cooling in a magnetic trap. However, thulium has been successfully captured in a magneto-optical trap (MOT) using a strong transition at 410 nm and a secondary cooling transition at 530 nm \cite{Sukachev2010,Sukachev2014,Kalganova2017}. The lifetime of the ${}^2F_{5/2}$ state was found to be about 75 ms in solid helium \cite{Ishikawa1997} and 112 ms in an optical lattice \cite{Sukachev2016}. Because of the small sensitivity to blackbody radiation, this transition is suggested as a possible optical clock \cite{Sukachev2016,Vishnyakova2016}.

\section{Experiment}

\begin{figure}
    \centering
    \includegraphics[width=1.0\linewidth]{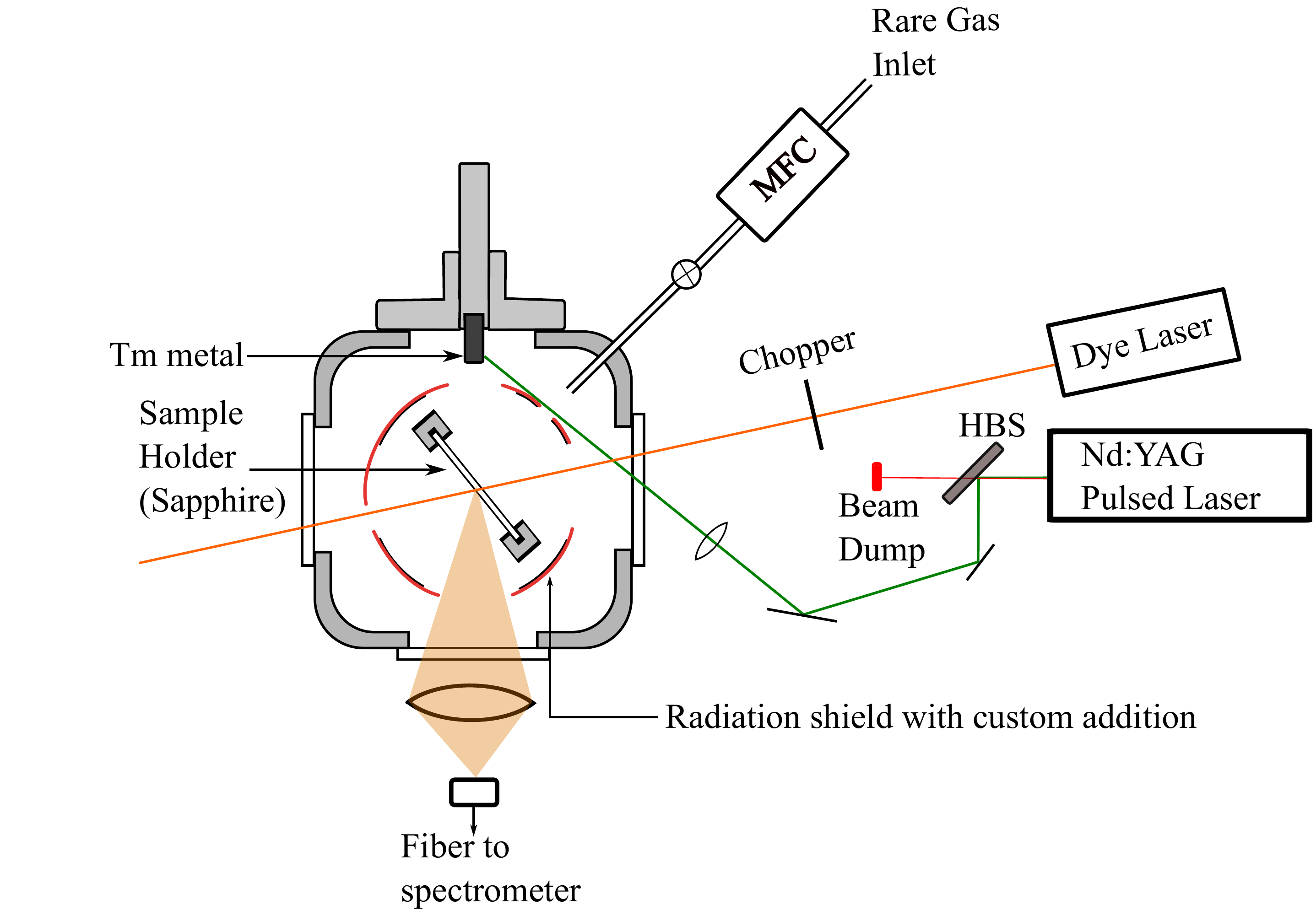}
    \caption{Experimental setup for trapping thulium atoms in rare gas matrices. The  flow of rare gas is controlled using the mass flow controller (MFC) and thulium atoms are obtained by ablation with a 532 nm pulsed laser (green beam).
    }
    \label{fig:schematics}
\end{figure}
The layout of our apparatus is shown in Fig. \ref{fig:schematics}. During the experiment,  the samples are prepared in a closed cycle helium cryostat with the cold head mounted vertically inside the vacuum shroud. The vacuum shroud is connected to the cryocooler using double O-rings, which help maintain vacuum inside the cryostat  while allowing rotation of the vacuum shroud. Prior to the growth, a dry scroll pump is used to maintain vacuum inside the cryostat. The thulium metal inside the chamber is mounted on an aluminum holder and kept under vacuum to prevent oxidation. 

A radiation shield is attached to the first stage of the cold head. The second stage heat station reaches the lowest temperature and the sample holder is attached to this stage with an indium gasket to maintain thermal contact. The sample holder is made of nickel plated copper and it holds a 19 mm diameter sapphire window for growing the argon/neon matrix. A silicon diode temperature sensor anchored to the sample holder is used to monitor and stabilize the temperature via a temperature controller and a resistive heater.

With the commercial radiation shield attached to the first stage heat station, a temperature of 15 K at the sample holder was achieved, which was not low enough to form crystals of neon. This is due to a higher temperature on the sapphire window, and a decrease of the 24.5 K freezing point at low pressure.  After using an additional radiation shield, we achieved a temperature of 11.5 K at the sample holder.

The rare gas to be deposited flows into the cryostat through a 1/16" outer diameter stainless steel tube entering through a connector on the vacuum shroud. Flow rate of the gases is in the range of 2 to 10 standard cm$^3$/min controlled by a mass flow controller. After a crystal of argon or neon is grown up to a certain thickness (usually 40-50 $\mu $m), the thulium growth is started by performing laser ablation with a frequency doubled  Q-switched  Nd:YAG pulse laser with a pulse width of 4-6 ns, a repetition rate of 20 Hz, and a pulse energy of 4 - 10 mJ. A harmonic beam splitter (HBS) separates the output of the laser insuring only second harmonic light at 532 nm reaches the sample. The thulium target (99.99 \% pure 1 g Tm)  is held in place by an aluminum rod passing through a vacuum fitting such as to allow manual rotation from the outside. The pulsed laser is focused  by a lens (f = 100 mm)  on a translatable stage to a  $ 10~ \mu m$ diameter spot on the thulium metal target with a fluence of about 10$^4$ J/cm$^2$ per pulse. During the ablation atoms and possibly some clusters of atoms are generated and deposited on the sample holder and get trapped into the crystal of argon or neon. Rare gas crystal growth is monitored by thin film interference using a laser diode and the deposition of thulium atoms in the crystal is monitored by absorption spectroscopy and laser induced fluorescence. The fluorescence signal in the near infrared region was detected with a monochromator and a InGaAs photodiode. A 50 mm diameter lens approximately 100 mm from the sample is used to collect the fluorescence signal, which is then focused by a small aspheric lens into a multimode optical fiber which runs to the monochromator. To improve the signal to noise ratio, we chopped the excitation beam and used a lock-in amplifier, enabling sensitive measurements of photocurrents as low as 100 fA.

\section{Results}
 
 % -------------- Two D spectra ---------------------
% -------------------------------------------------------
  \begin{figure*}
  \centering
  \includegraphics[width=1.0\linewidth]{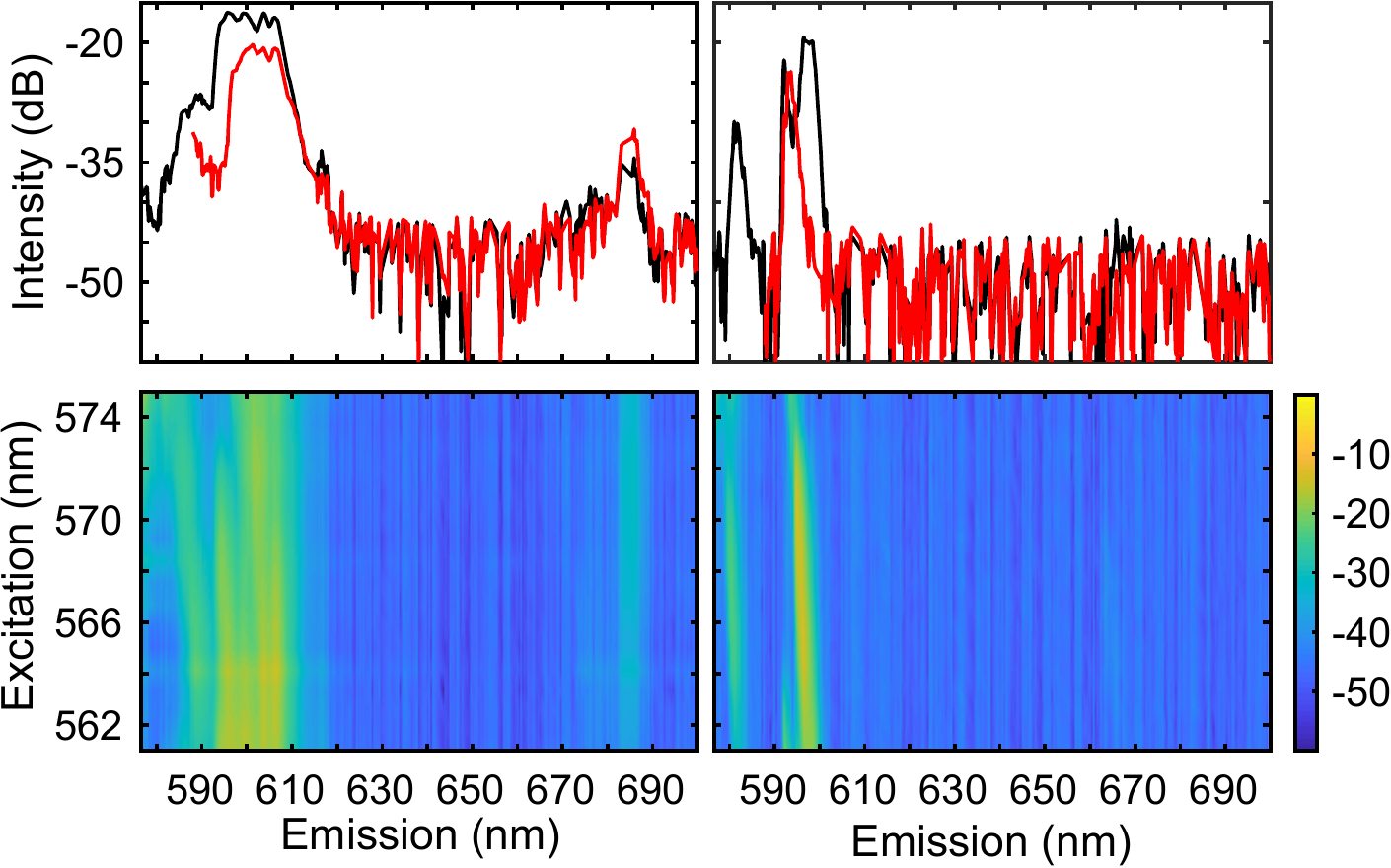}
\caption{Two dimensional emission  spectra of thulium in argon (left) and neon (right) at different excitation wavelengths normalized by excitation power. The top two figures are the observed fluorescence signal at excitation wavelengths 561 nm (black) and 580.6 nm (red) curves.}
\label{fig:2Dspectra}
\end{figure*}
% -------------- Two D spectra ---------------------
% -------------------------------------------------------
 
\begin{figure}
    \centering
    \includegraphics[width=1.0\linewidth]{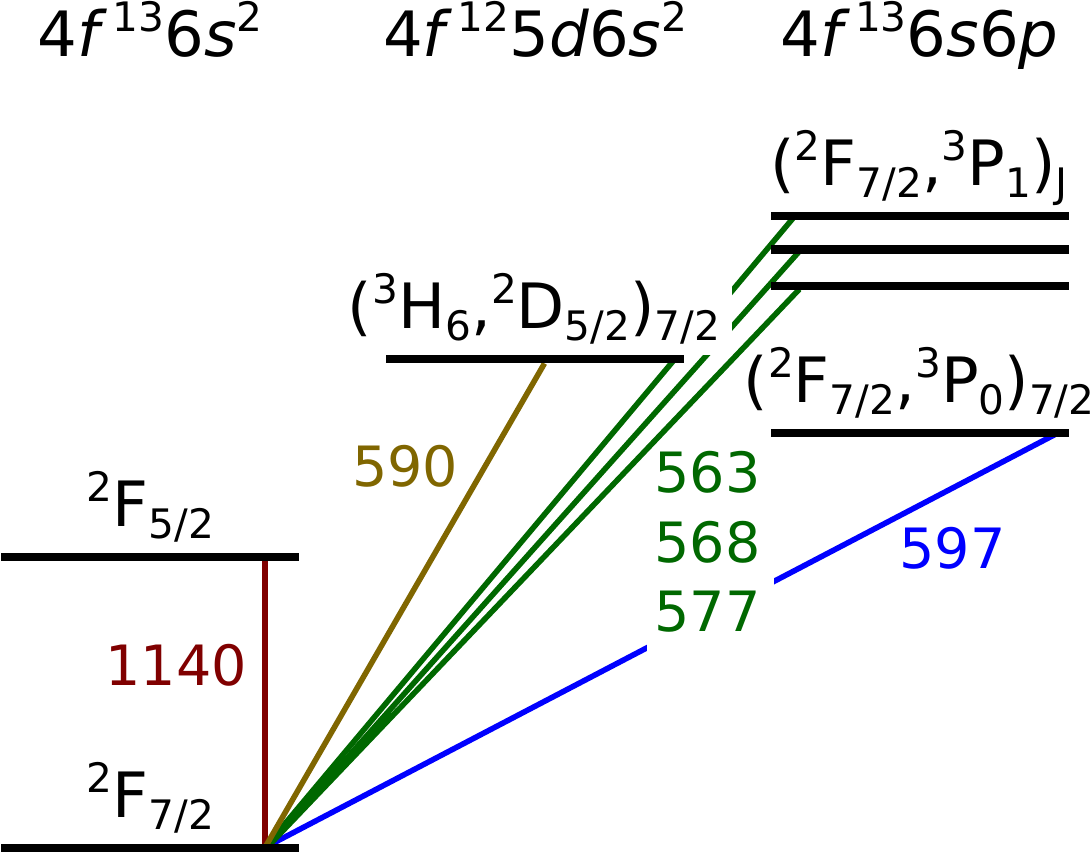}
    \caption{Diagram of the relevant energy levels for thulium atom in vacuum. The numbers on the transitions indicate wavelength in nm.
    }
    \label{fig:level_diagram}
\end{figure}

In order to characterize the emission properties of thulium in argon/neon crystals, we performed laser induced fluorescence spectroscopy. The sample was excited with a dye laser which was tuned from 562 nm to 585 nm while we recorded the excitation-emission spectra using a spectrometer as shown in Fig. \ref{fig:2Dspectra}. Typical excitation power was 10 mW. To determine the optimal gas flow rate and ablation laser power we performed spectroscopy under a variety of growth conditions.  The overall intensity of the spectra depended on these conditions; however, the positions  of the features remained roughly the same. Under optimal growth conditions, we deposit a crystal that is approximately 130 $\mu\textrm{m}$ thick in one hour. Based on the film thickness of samples grown without a noble gas matrix, we estimate the concentration of thulium to be of the order of several parts per thousand. The two dimensional spectra show strong fluorescence features in the range of 580-600 nm when the excitation is around 560-580 nm. There are 5 known transitions involving the ground state for thulium atoms in vacuum in this wavelength range: three lines at 563, 568, and 577 nm involving the $4f^{13}6s6p (7/2,1)_J$ levels, a line at 590 nm involving the $4f^{12}5d6s^2 (6,5/2)_{7/2}$ level, and a line at 597 nm involving the $4f^{13}6s6p (7/2,0)_{7/2}$ level\cite{NIST_ASD}. Figure \ref{fig:level_diagram} summarizes the relevant states for thulium in vacuum.

It is difficult to make specific assignments to these levels in argon, as they could be mixed by interaction with the host. In neon the strongest fluorescence is in the range 593-600 nm. Depending on the excitation wavelength, the emission band shifts closer to 593 nm or 600 nm, suggesting that these features have different origins. The sharper 593 nm emission band has been attributed previously to the vacuum 590 nm line\cite{Belyaeva1986}, as the narrower width for 593 nm emission is consistent with an inner shell $f$-$d$ excitation. We assign the 600 nm emission band to the vacuum 597 nm line. This is further supported by the dependence on excitation wavelength, which matches the pattern of a weaker emission band at 582 nm that can be assigned to the vacuum 577 nm line from the same configuration. In this picture, different excitation wavelengths preferentially populate the $4f^{13}6s6p$ or $4f^{12}5d6s^2$ configurations, and non-radiative relaxation to the lowest state within the $4f^{12}6s6p$ configuration leads to a dominant emission band at 600 nm. There is an additional weak feature in argon at 685 nm, which is not clearly associated with any known transitions in free Tm atoms. It cannot be excluded that clusters or oxides of thulium could be contributing to these observations, however, the strength of the fluorescence features correlates with our subsequent observation of infrared lines at 1140 nm, meaning the observed features are likely to be coming from atomic thulium.
% -------------------------------------------------------

% ---------------------------------------------------------------
\begin{figure}
\centering
 \includegraphics[width=1\linewidth]{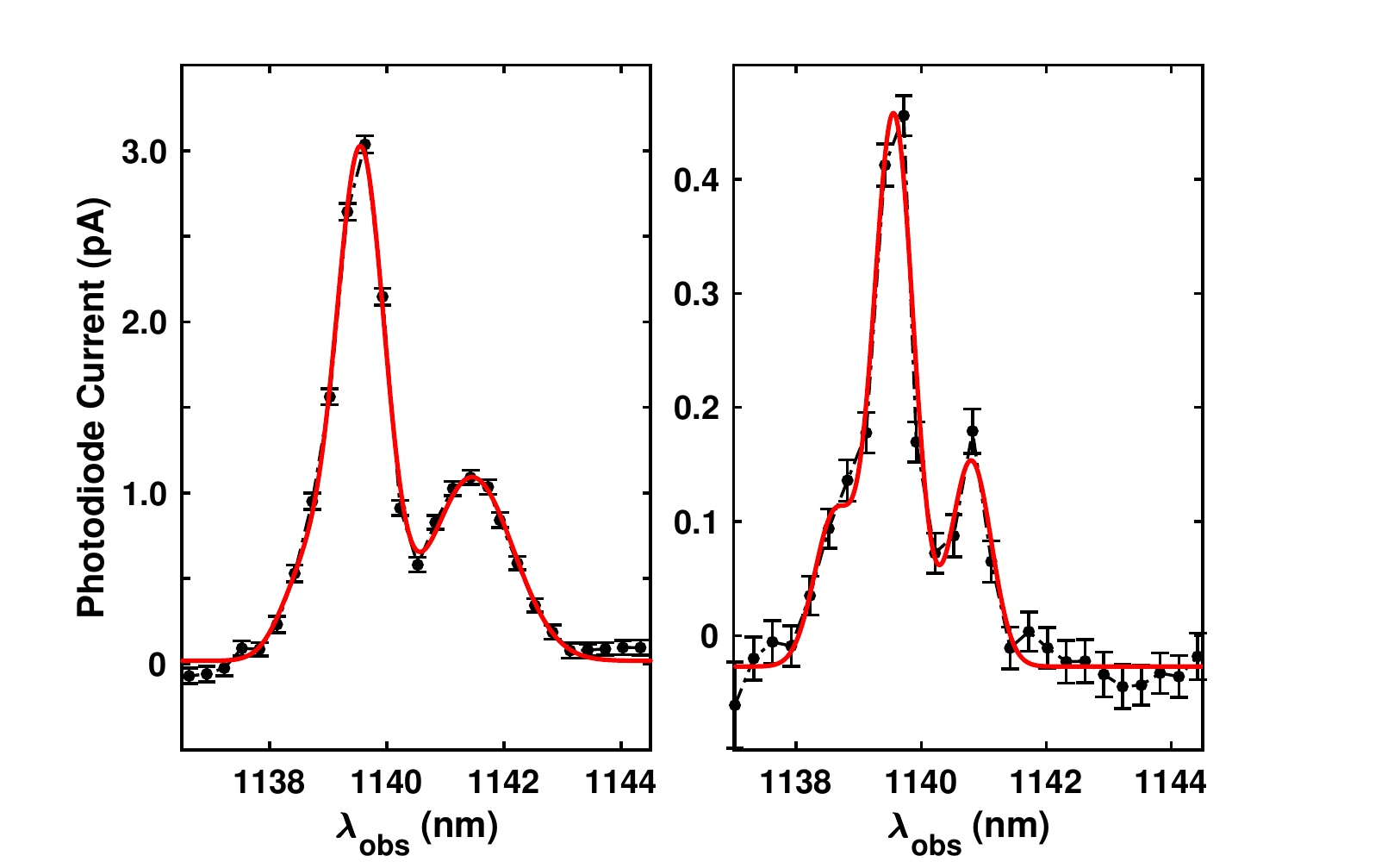}
 \caption{The emission spectra of the ${}^2F_{5/2} - {}^2F_{7/2}$ transition in argon (left) and in neon (right) with excitation at 594.8 nm. The solid red curves are fits to a model based on three resolution-limited gaussian curves (full width and half maximum: 0.8 nm). Error bars indicate statistical uncertainty.}
  \label{fig:1140nm}
\end{figure}
%-----------------------------------------------------------------------
 \begin{figure}
  \centering
  \includegraphics[width=1.0\linewidth]{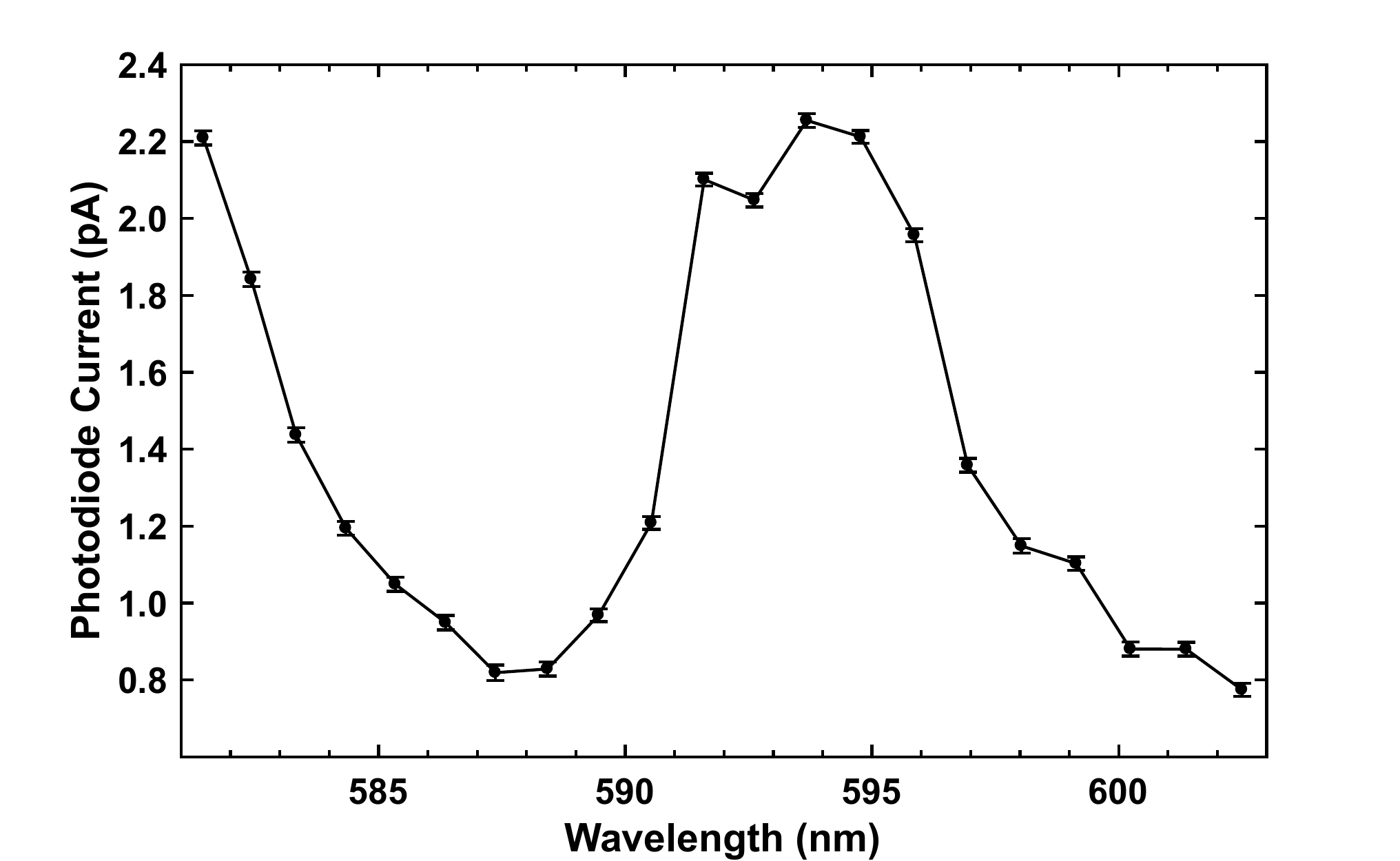}
\caption{Excitation spectrum of thulium measured in a neon crystal from 582 to 602 nm. The monitored wavelength was 1140 nm. Error bars indicate statistical uncertainty. The excitation laser linewidth is less than 0.1 nm.}
\label{fig:excitationSpectrum}
\end{figure}
% ---------------------------------------------------------------
% Table for 1140 line
\begin{table}
\caption{Fitted spectral peaks}
\begin{tabular}{ |c|c c|}
 \hline
 \hline
  & \multicolumn{2}{c|}{Peak Current (pA)}\\
 \cline{2-3}
  Peak Position (nm) &~~ Argon & Neon\\
 \hline
 1138.6 & ~0.48 & 0.13\\
 \hline
 1139.6 & ~2.94 & 0.48\\
 \hline
 1140.8 & ~- & 0.18 \\
 \hline
 1141.5 & ~1.07 &-\\
 \hline 
\end{tabular}
\label{table:1140Line}
\end{table}

Figure \ref{fig:1140nm} shows our main result, the narrow emission lines of thulium near 1140 nm in both argon and neon corresponding with the magnetic dipole transition from ${}^2F_{5/2} $ to ${}^2F_{7/2}$ states in free atoms. The observed width of these features is consistent with our monochromator limit determined at 1160 nm using the second diffracted order of the dye laser. The laser spectral purity and absolute wavelength is independently verified with a high resolution wavelength meter. This transition is shielded by the outer electrons and is less affected by the presence of the argon or neon so the shift of this emission is small compared with that seen in the visible region. However, the transition is split, with a central feature and a clear satellite feature on the red side. Another weaker and poorly resolved feature is likely present on the blue side. A fit to the features using three Gaussian peaks with the monochromator-limited width yields the positions and intensities shown in table \ref{table:1140Line}. One possible origin of the splitting, which was not observed in solid helium, is the presence of a crystal field. This effect arises from the breaking of the full rotational symmetry of a free atom into the discrete point group symmetry of the crystal. This effect is known to be much larger (10s of nm) for rare earth atoms trapped in fluoride crystals\cite{Miller:1970hrba,Bespalov:1997kzba,Sugiyama:2006chba}, however, the much weaker van der Waals bonding in noble gas solids might contribute a scaled down version of this effect. It is also possible that the splitting could be due to multiple trapping sites with a small shift between them, but this would have to occur similarly in both neon and argon. The splitting and the relative intensities of the peaks were consistent across multiple growths.

If the observed splitting is indeed due to crystal field effects, it is worth considering relevant symmetries. For fluoride crystals, the trapping site can have the cubic (also known as full octahedral, $O_h$) symmetry or a lower symmetry. In the former case the ${}^2F_{7/2}$ level splits into 3 sub-levels, and the ${}^2F_{5/2}$ splits into 2 sublevels. In the latter case the levels are maximally split into 3 and 4 sub-levels, respectively\cite{Bespalov:1997kzba}. Since both neon and argon have fcc crystal structures, it is possible that the trapping center could fall into either category. In either case more than 3 lines are expected, but some may be too weak or close together to resolve at present. Thulium atoms trapped in helium had no observed splitting\cite{Ishikawa1997}, but lighter and less tightly bound helium atoms would be expected to move around and present less rotational symmetry breaking.

In helium, zero-phonon features were observed as a sharp peak in the fluorescence as a function of the excitation wavelength at 591 nm\cite{Ishikawa1997}. The dependence of the narrow 1140 nm emission intensity on excitation wavelength for neon is shown in the relevant range in Fig. \ref{fig:excitationSpectrum}. There is strong fluorescence for all blue wavelengths, which begins to fall above 580 nm. In addition, there is a peak near 594 nm, likely an excitation of the $4f^{13}6s6p (7/2,0)_{7/2}$ level corresponding to the vacuum 597 nm line. It cannot be excluded that near 591 nm there is a small bump due to a zero-phonon peak partially masked by the 594 nm peak, but this feature is not reproduced in all growths. It is not surprising that zero-phonon features should be weaker in heavier host crystals, as the heavier the host the less the host atom configurations between the ground and excited states will overlap due to zero-point motion.

Over several hours of measurements, the signal strength decreases and there appears a bleached spot on the sample where the dye laser illuminated it. Ordinarily the thulium-doped crystal has a brown appearance, but it is closer to transparent or white in the bleached area. Exposing the sample to white light and/or blocking the excitation beam does not cause the bleached spot to recover. However, annealing the sample partially recovers the fluorescence. Typically the bleached spot is several millimeters across, consistent with the illuminated area size. It is difficult to explain this effect by either clustering or diffusion, since clustering would presumably not be reversible, and diffusion would reach the edge fastest along the direction in which the sample is thinnest and either coat the substrate or form a metallic layer at the free surface, neither of which is observed. Thus the nature of this degradation is not fully understood. However, the spectra are acquired sufficiently quickly (10-30 minutes) that the degradation between the beginning and end of a spectrum is less than 10\%. No difference in the spectra other than an overall decrease is seen, although the signal-to-noise after bleaching is unfavorable for detailed comparison. Bleaching effects are not uncommon in matrix isolation and have been seen in Rb in noble gases\cite{Ilja:2012,Kanagin2013} and parahydrogen\cite{upadhyay2016}.

%------------------------------------------------------------

% ---------------------------------------------------------------

\begin{figure}
\centering
\includegraphics[width=1.1\linewidth]{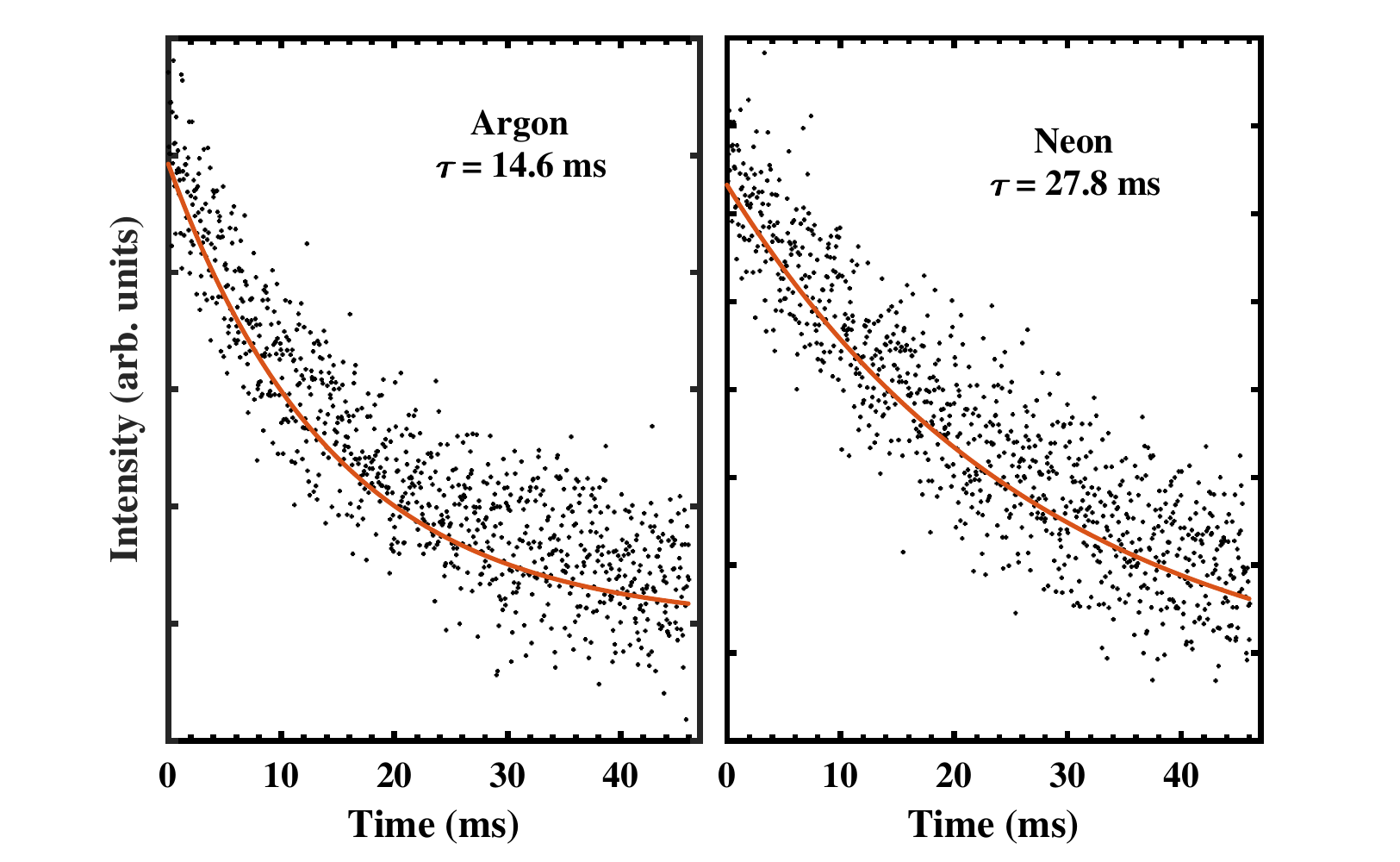}
\caption{The decay of the ${}^2F_{5/2} - {}^2F_{7/2}$ transition  of thulium in argon and  neon. The solid line is obtained by fitting a curve of the form $A ~e^{-t/\tau}+B$, with the lifetime $\tau$ indicated.}
\label{fig:Lifetimes}
\end{figure}

We measured the lifetime of the metastable ${}^2F_{5/2} $ state by exciting the sample with a periodic series of exciting pulses by chopping the dye laser beam at a constant frequency and recording the fluorescence signal using a current amplifier and oscilloscope. This measurement technique gives a lower signal-to-noise ratio than the lock-in, and so each lifetime trace requires about 90 minutes to acquire. Figure \ref{fig:Lifetimes} shows the decay of emission intensity of the 1140 nm line in argon and neon. The lifetimes are calculated by fitting an exponential function of the form $A ~e^{-t/\tau}+B$  to the data, where $\tau$ is the lifetime of the ${}^2F_{5/2} $ state. The measured values were 14.6(0.5) ms in argon and 28(3) ms in neon. The quoted uncertainties represent the effects of statistical errors on the fit, assuming uncorrelated noise. Both of these values are much smaller than the lifetime of 75(3) ms measured in solid helium \cite{Ishikawa1997} and 112(4) ms in a 1D-optical lattice \cite{Sukachev2016}. We speculate that this might be explained by a non-centrosymmetric trapping site leading to a small electric dipole matrix element.

% -------------------------------------------------------
% -------------- Fluorescence Decay  ---------------------
% -------------------------------------------------------

\section{Discussion}
For surface sensing applications, thulium trapped in argon or neon crystals appears to be a promising candidate. As host crystals, compared with helium, these are easier to deposit, and higher concentrations of thulium are possible. Parahydrogen and orthodeuterium are other potential candidates, although these are solid over a lower range of temperatures. From the atomic perspective, the heavier noble gases seem to introduce additional effects, such as a splitting, and reduce slightly the lifetime of the excited state. Neither of these is likely to impact the suitably for surface sensing, and in fact, if the lifetime reduction comes from a larger optical transition matrix element, this would be beneficial for applications where the 1140 nm transition was used for excitation. Importantly, the linewidth appears to remain narrow, and what the true linewidth is remains an open question. Already, the linewidth is at the level where large Zeeman shifts from Tesla-scale fields can be detected\cite{Ishikawa1997}. If the true linewidth is significantly narrower, correspondingly smaller fields could be seen and one might imagine detecting, for example, nanoscale field inhomogeneity in a thin layer above an antiferromagnet or an exotic spin system. If the hyperfine coupling can be resolved, optical pumping might be employed to use the magnetic sublevels as an even more sensitive magnetometer. Finally, we note that if the line splitting is indeed due to crystal field effects, there exists the possibility to do optical refrigeration\cite{Epstein1995,Sheik-Bahae2016}, on a much smaller energy and temperature scale, provided the non-radiative and Raman processes can be kept small. 

\section{Conclusion}
We prepared samples of thulium atoms trapped in a crystals of argon and neon.  We performed visible and infrared laser-induced fluorescence spectroscopy. The spectra were shifted and broadened from the vapor phase in the visible range, but in the near-infrared region, the emission spectrum remains narrow and splits into multiple lines. These results suggest that this narrow optical transition occurs between two levels with nearly identical coupling to the surrounding environment. Direct excitation of this transition could allow for optical pumping, and possibly even optical refrigeration. To determine the true linewidth we plan to improve our spectroscopy resolution in the near future.


\begin{thebibliography}{39}%
\makeatletter
\providecommand \@ifxundefined [1]{%
 \@ifx{#1\undefined}
}%
\providecommand \@ifnum [1]{%
 \ifnum #1\expandafter \@firstoftwo
 \else \expandafter \@secondoftwo
 \fi
}%
\providecommand \@ifx [1]{%
 \ifx #1\expandafter \@firstoftwo
 \else \expandafter \@secondoftwo
 \fi
}%
\providecommand \natexlab [1]{#1}%
\providecommand \enquote  [1]{``#1''}%
\providecommand \bibnamefont  [1]{#1}%
\providecommand \bibfnamefont [1]{#1}%
\providecommand \citenamefont [1]{#1}%
\providecommand \href@noop [0]{\@secondoftwo}%
\providecommand \href [0]{\begingroup \@sanitize@url \@href}%
\providecommand \@href[1]{\@@startlink{#1}\@@href}%
\providecommand \@@href[1]{\endgroup#1\@@endlink}%
\providecommand \@sanitize@url [0]{\catcode `\\12\catcode `\$12\catcode
  `\&12\catcode `\#12\catcode `\^12\catcode `\_12\catcode `\%12\relax}%
\providecommand \@@startlink[1]{}%
\providecommand \@@endlink[0]{}%
\providecommand \url  [0]{\begingroup\@sanitize@url \@url }%
\providecommand \@url [1]{\endgroup\@href {#1}{\urlprefix }}%
\providecommand \urlprefix  [0]{URL }%
\providecommand \Eprint [0]{\href }%
\providecommand \doibase [0]{http://dx.doi.org/}%
\providecommand \selectlanguage [0]{\@gobble}%
\providecommand \bibinfo  [0]{\@secondoftwo}%
\providecommand \bibfield  [0]{\@secondoftwo}%
\providecommand \translation [1]{[#1]}%
\providecommand \BibitemOpen [0]{}%
\providecommand \bibitemStop [0]{}%
\providecommand \bibitemNoStop [0]{.\EOS\space}%
\providecommand \EOS [0]{\spacefactor3000\relax}%
\providecommand \BibitemShut  [1]{\csname bibitem#1\endcsname}%
\let\auto@bib@innerbib\@empty
%</preamble>
\bibitem [{\citenamefont {Helmcke}\ \emph {et~al.}(2003)\citenamefont
  {Helmcke}, \citenamefont {Wilpers}, \citenamefont {Binnewies}, \citenamefont
  {Degenhardt}, \citenamefont {Sterr}, \citenamefont {Schnatz},\ and\
  \citenamefont {Riehle}}]{Helmcke2003}%
  \BibitemOpen
  \bibfield  {author} {\bibinfo {author} {\bibfnamefont {J.}~\bibnamefont
  {Helmcke}}, \bibinfo {author} {\bibfnamefont {G.}~\bibnamefont {Wilpers}},
  \bibinfo {author} {\bibfnamefont {T.}~\bibnamefont {Binnewies}}, \bibinfo
  {author} {\bibfnamefont {C.}~\bibnamefont {Degenhardt}}, \bibinfo {author}
  {\bibfnamefont {U.}~\bibnamefont {Sterr}}, \bibinfo {author} {\bibfnamefont
  {H.}~\bibnamefont {Schnatz}}, \ and\ \bibinfo {author} {\bibfnamefont
  {F.}~\bibnamefont {Riehle}},\ }\href {\doibase 10.1109/TIM.2003.810025}
  {\bibfield  {journal} {\bibinfo  {journal} {IEEE Transactions on
  Instrumentation and Measurement}\ }\textbf {\bibinfo {volume} {52}},\
  \bibinfo {pages} {250} (\bibinfo {year} {2003})}\BibitemShut {NoStop}%
\bibitem [{\citenamefont {Ushijima}\ \emph {et~al.}(2015)\citenamefont
  {Ushijima}, \citenamefont {Takamoto}, \citenamefont {Das}, \citenamefont
  {Ohkubo},\ and\ \citenamefont {Katori}}]{Ushijima2015}%
  \BibitemOpen
  \bibfield  {author} {\bibinfo {author} {\bibfnamefont {I.}~\bibnamefont
  {Ushijima}}, \bibinfo {author} {\bibfnamefont {M.}~\bibnamefont {Takamoto}},
  \bibinfo {author} {\bibfnamefont {M.}~\bibnamefont {Das}}, \bibinfo {author}
  {\bibfnamefont {T.}~\bibnamefont {Ohkubo}}, \ and\ \bibinfo {author}
  {\bibfnamefont {H.}~\bibnamefont {Katori}},\ }\href
  {https://doi.org/10.1038/nphoton.2015.5} {\bibfield  {journal} {\bibinfo
  {journal} {Nature Photonics}\ }\textbf {\bibinfo {volume} {9}},\ \bibinfo
  {pages} {185 EP} (\bibinfo {year} {2015})}\BibitemShut {NoStop}%
\bibitem [{\citenamefont {Ludlow}\ \emph {et~al.}(2015)\citenamefont {Ludlow},
  \citenamefont {Boyd}, \citenamefont {Ye}, \citenamefont {Peik},\ and\
  \citenamefont {Schmidt}}]{RevModPhys.87.637}%
  \BibitemOpen
  \bibfield  {author} {\bibinfo {author} {\bibfnamefont {A.~D.}\ \bibnamefont
  {Ludlow}}, \bibinfo {author} {\bibfnamefont {M.~M.}\ \bibnamefont {Boyd}},
  \bibinfo {author} {\bibfnamefont {J.}~\bibnamefont {Ye}}, \bibinfo {author}
  {\bibfnamefont {E.}~\bibnamefont {Peik}}, \ and\ \bibinfo {author}
  {\bibfnamefont {P.~O.}\ \bibnamefont {Schmidt}},\ }\href {\doibase
  10.1103/RevModPhys.87.637} {\bibfield  {journal} {\bibinfo  {journal} {Rev.
  Mod. Phys.}\ }\textbf {\bibinfo {volume} {87}},\ \bibinfo {pages} {637}
  (\bibinfo {year} {2015})}\BibitemShut {NoStop}%
\bibitem [{\citenamefont {Budker}\ and\ \citenamefont
  {Romalis}(2007)}]{Budker2007}%
  \BibitemOpen
  \bibfield  {author} {\bibinfo {author} {\bibfnamefont {D.}~\bibnamefont
  {Budker}}\ and\ \bibinfo {author} {\bibfnamefont {M.}~\bibnamefont
  {Romalis}},\ }\href {https://doi.org/10.1038/nphys566} {\bibfield  {journal}
  {\bibinfo  {journal} {Nature Physics}\ }\textbf {\bibinfo {volume} {3}},\
  \bibinfo {pages} {227 EP } (\bibinfo {year} {2007})}\BibitemShut {NoStop}%
\bibitem [{\citenamefont {Ishikawa}\ \emph {et~al.}(1997)\citenamefont
  {Ishikawa}, \citenamefont {Hatakeyama}, \citenamefont {Gosyono-o},
  \citenamefont {Wada}, \citenamefont {Takahashi},\ and\ \citenamefont
  {Yabuzaki}}]{Ishikawa1997}%
  \BibitemOpen
  \bibfield  {author} {\bibinfo {author} {\bibfnamefont {K.}~\bibnamefont
  {Ishikawa}}, \bibinfo {author} {\bibfnamefont {A.}~\bibnamefont
  {Hatakeyama}}, \bibinfo {author} {\bibfnamefont {K.}~\bibnamefont
  {Gosyono-o}}, \bibinfo {author} {\bibfnamefont {S.}~\bibnamefont {Wada}},
  \bibinfo {author} {\bibfnamefont {Y.}~\bibnamefont {Takahashi}}, \ and\
  \bibinfo {author} {\bibfnamefont {T.}~\bibnamefont {Yabuzaki}},\ }\href
  {\doibase 10.1103/PhysRevB.56.780} {\bibfield  {journal} {\bibinfo  {journal}
  {Phys. Rev. B}\ }\textbf {\bibinfo {volume} {56}},\ \bibinfo {pages} {780}
  (\bibinfo {year} {1997})}\BibitemShut {NoStop}%
\bibitem [{\citenamefont {de~Riedmatten}\ \emph {et~al.}(2008)\citenamefont
  {de~Riedmatten}, \citenamefont {Afzelius}, \citenamefont {Staudt},
  \citenamefont {Simon},\ and\ \citenamefont {Gisin}}]{Riedmatten2008}%
  \BibitemOpen
  \bibfield  {author} {\bibinfo {author} {\bibfnamefont {H.}~\bibnamefont
  {de~Riedmatten}}, \bibinfo {author} {\bibfnamefont {M.}~\bibnamefont
  {Afzelius}}, \bibinfo {author} {\bibfnamefont {M.~U.}\ \bibnamefont
  {Staudt}}, \bibinfo {author} {\bibfnamefont {C.}~\bibnamefont {Simon}}, \
  and\ \bibinfo {author} {\bibfnamefont {N.}~\bibnamefont {Gisin}},\
  }\href@noop {} {\bibfield  {journal} {\bibinfo  {journal} {Nature}\ }\textbf
  {\bibinfo {volume} {456}},\ \bibinfo {pages} {773} (\bibinfo {year}
  {2008})}\BibitemShut {NoStop}%
\bibitem [{\citenamefont {Robledo}\ \emph {et~al.}(2011)\citenamefont
  {Robledo}, \citenamefont {Childress}, \citenamefont {Bernien}, \citenamefont
  {Hensen}, \citenamefont {Alkemade},\ and\ \citenamefont
  {Hanson}}]{Robledo2011}%
  \BibitemOpen
  \bibfield  {author} {\bibinfo {author} {\bibfnamefont {L.}~\bibnamefont
  {Robledo}}, \bibinfo {author} {\bibfnamefont {L.}~\bibnamefont {Childress}},
  \bibinfo {author} {\bibfnamefont {H.}~\bibnamefont {Bernien}}, \bibinfo
  {author} {\bibfnamefont {B.}~\bibnamefont {Hensen}}, \bibinfo {author}
  {\bibfnamefont {P.~F.~A.}\ \bibnamefont {Alkemade}}, \ and\ \bibinfo {author}
  {\bibfnamefont {R.}~\bibnamefont {Hanson}},\ }\href@noop {} {\bibfield
  {journal} {\bibinfo  {journal} {Nature}\ }\textbf {\bibinfo {volume} {477}},\
  \bibinfo {pages} {574} (\bibinfo {year} {2011})}\BibitemShut {NoStop}%
\bibitem [{\citenamefont {Raston}\ \emph {et~al.}(2013)\citenamefont {Raston},
  \citenamefont {Kettwich},\ and\ \citenamefont {Anderson}}]{Raston:2013}%
  \BibitemOpen
  \bibfield  {author} {\bibinfo {author} {\bibfnamefont {P.~L.}\ \bibnamefont
  {Raston}}, \bibinfo {author} {\bibfnamefont {S.~C.}\ \bibnamefont
  {Kettwich}}, \ and\ \bibinfo {author} {\bibfnamefont {D.~T.}\ \bibnamefont
  {Anderson}},\ }\href {https://doi.org/10.1063/1.4820528} {\bibfield
  {journal} {\bibinfo  {journal} {The Journal of Chemical Physics}\ }\textbf
  {\bibinfo {volume} {139}},\ \bibinfo {pages} {134304} (\bibinfo {year}
  {2013})}\BibitemShut {NoStop}%
\bibitem [{\citenamefont {Parker}\ \emph {et~al.}(2010)\citenamefont {Parker},
  \citenamefont {Aynajian}, \citenamefont {da~Silva~Neto}, \citenamefont
  {Pushp}, \citenamefont {Ono}, \citenamefont {Wen}, \citenamefont {Xu},
  \citenamefont {Gu},\ and\ \citenamefont {Yazdani}}]{Parker:2010}%
  \BibitemOpen
  \bibfield  {author} {\bibinfo {author} {\bibfnamefont {C.~V.}\ \bibnamefont
  {Parker}}, \bibinfo {author} {\bibfnamefont {P.}~\bibnamefont {Aynajian}},
  \bibinfo {author} {\bibfnamefont {E.}~\bibnamefont {da~Silva~Neto}}, \bibinfo
  {author} {\bibfnamefont {A.}~\bibnamefont {Pushp}}, \bibinfo {author}
  {\bibfnamefont {S.}~\bibnamefont {Ono}}, \bibinfo {author} {\bibfnamefont
  {J.}~\bibnamefont {Wen}}, \bibinfo {author} {\bibfnamefont {Z.}~\bibnamefont
  {Xu}}, \bibinfo {author} {\bibfnamefont {G.}~\bibnamefont {Gu}}, \ and\
  \bibinfo {author} {\bibfnamefont {A.}~\bibnamefont {Yazdani}},\ }\href@noop
  {} {\bibfield  {journal} {\bibinfo  {journal} {Nature}\ }\textbf {\bibinfo
  {volume} {468}},\ \bibinfo {pages} {677} (\bibinfo {year}
  {2010})}\BibitemShut {NoStop}%
\bibitem [{\citenamefont {Comin}\ \emph {et~al.}(2014)\citenamefont {Comin},
  \citenamefont {Frano}, \citenamefont {Yee}, \citenamefont {Yoshida},
  \citenamefont {Eisaki}, \citenamefont {Schierle}, \citenamefont {Weschke},
  \citenamefont {Sutarto}, \citenamefont {He}, \citenamefont {Soumyanarayanan},
  \citenamefont {He}, \citenamefont {Le~Tacon}, \citenamefont {Elfimov},
  \citenamefont {Hoffman}, \citenamefont {Sawatzky}, \citenamefont {Keimer},\
  and\ \citenamefont {Damascelli}}]{Comin:2014}%
  \BibitemOpen
  \bibfield  {author} {\bibinfo {author} {\bibfnamefont {R.}~\bibnamefont
  {Comin}}, \bibinfo {author} {\bibfnamefont {A.}~\bibnamefont {Frano}},
  \bibinfo {author} {\bibfnamefont {M.~M.}\ \bibnamefont {Yee}}, \bibinfo
  {author} {\bibfnamefont {Y.}~\bibnamefont {Yoshida}}, \bibinfo {author}
  {\bibfnamefont {H.}~\bibnamefont {Eisaki}}, \bibinfo {author} {\bibfnamefont
  {E.}~\bibnamefont {Schierle}}, \bibinfo {author} {\bibfnamefont
  {E.}~\bibnamefont {Weschke}}, \bibinfo {author} {\bibfnamefont
  {R.}~\bibnamefont {Sutarto}}, \bibinfo {author} {\bibfnamefont
  {F.}~\bibnamefont {He}}, \bibinfo {author} {\bibfnamefont {A.}~\bibnamefont
  {Soumyanarayanan}}, \bibinfo {author} {\bibfnamefont {Y.}~\bibnamefont {He}},
  \bibinfo {author} {\bibfnamefont {M.}~\bibnamefont {Le~Tacon}}, \bibinfo
  {author} {\bibfnamefont {I.~S.}\ \bibnamefont {Elfimov}}, \bibinfo {author}
  {\bibfnamefont {J.~E.}\ \bibnamefont {Hoffman}}, \bibinfo {author}
  {\bibfnamefont {G.~A.}\ \bibnamefont {Sawatzky}}, \bibinfo {author}
  {\bibfnamefont {B.}~\bibnamefont {Keimer}}, \ and\ \bibinfo {author}
  {\bibfnamefont {A.}~\bibnamefont {Damascelli}},\ }\href {\doibase
  10.1126/science.1242996} {\bibfield  {journal} {\bibinfo  {journal}
  {Science}\ }\textbf {\bibinfo {volume} {343}},\ \bibinfo {pages} {390}
  (\bibinfo {year} {2014})},\ \Eprint
  {http://arxiv.org/abs/http://science.sciencemag.org/content/343/6169/390.full.pdf}
  {http://science.sciencemag.org/content/343/6169/390.full.pdf} \BibitemShut
  {NoStop}%
\bibitem [{\citenamefont {Rosenthal}\ \emph {et~al.}(0209)\citenamefont
  {Rosenthal}, \citenamefont {Andrade}, \citenamefont {Arguello}, \citenamefont
  {Fernandes}, \citenamefont {Xing}, \citenamefont {Wang}, \citenamefont {Jin},
  \citenamefont {Millis},\ and\ \citenamefont {Pasupathy}}]{Rosenthal:2014}%
  \BibitemOpen
  \bibfield  {author} {\bibinfo {author} {\bibfnamefont {E.~P.}\ \bibnamefont
  {Rosenthal}}, \bibinfo {author} {\bibfnamefont {E.~F.}\ \bibnamefont
  {Andrade}}, \bibinfo {author} {\bibfnamefont {C.~J.}\ \bibnamefont
  {Arguello}}, \bibinfo {author} {\bibfnamefont {R.~M.}\ \bibnamefont
  {Fernandes}}, \bibinfo {author} {\bibfnamefont {L.~Y.}\ \bibnamefont {Xing}},
  \bibinfo {author} {\bibfnamefont {X.~C.}\ \bibnamefont {Wang}}, \bibinfo
  {author} {\bibfnamefont {C.~Q.}\ \bibnamefont {Jin}}, \bibinfo {author}
  {\bibfnamefont {A.~J.}\ \bibnamefont {Millis}}, \ and\ \bibinfo {author}
  {\bibfnamefont {A.~N.}\ \bibnamefont {Pasupathy}},\ }\href@noop {} {\bibfield
   {journal} {\bibinfo  {journal} {Nature Physics}\ }\textbf {\bibinfo {volume}
  {10}},\ \bibinfo {pages} {225} (\bibinfo {year} {20140209})}\BibitemShut
  {NoStop}%
\bibitem [{\citenamefont {Arguello}\ \emph {et~al.}(2015)\citenamefont
  {Arguello}, \citenamefont {Rosenthal}, \citenamefont {Andrade}, \citenamefont
  {Jin}, \citenamefont {Yeh}, \citenamefont {Zaki}, \citenamefont {Jia},
  \citenamefont {Cava}, \citenamefont {Fernandes}, \citenamefont {Millis},
  \citenamefont {Valla}, \citenamefont {R~M~Osgood},\ and\ \citenamefont
  {Pasupathy}}]{Arguello:2015}%
  \BibitemOpen
  \bibfield  {author} {\bibinfo {author} {\bibfnamefont {C.~J.}\ \bibnamefont
  {Arguello}}, \bibinfo {author} {\bibfnamefont {E.~P.}\ \bibnamefont
  {Rosenthal}}, \bibinfo {author} {\bibfnamefont {E.~F.}\ \bibnamefont
  {Andrade}}, \bibinfo {author} {\bibfnamefont {W.}~\bibnamefont {Jin}},
  \bibinfo {author} {\bibfnamefont {P.~C.}\ \bibnamefont {Yeh}}, \bibinfo
  {author} {\bibfnamefont {N.}~\bibnamefont {Zaki}}, \bibinfo {author}
  {\bibfnamefont {S.}~\bibnamefont {Jia}}, \bibinfo {author} {\bibfnamefont
  {R.~J.}\ \bibnamefont {Cava}}, \bibinfo {author} {\bibfnamefont {R.~M.}\
  \bibnamefont {Fernandes}}, \bibinfo {author} {\bibfnamefont {A.~J.}\
  \bibnamefont {Millis}}, \bibinfo {author} {\bibfnamefont {T.}~\bibnamefont
  {Valla}}, \bibinfo {author} {\bibfnamefont {J.}~\bibnamefont {R~M~Osgood}}, \
  and\ \bibinfo {author} {\bibfnamefont {A.~N.}\ \bibnamefont {Pasupathy}},\
  }\href@noop {} {\bibfield  {journal} {\bibinfo  {journal} {Physical Review
  Letters}\ }\textbf {\bibinfo {volume} {114}},\ \bibinfo {pages} {037001}
  (\bibinfo {year} {2015})}\BibitemShut {NoStop}%
\bibitem [{\citenamefont {Savary}\ and\ \citenamefont
  {Balents}(2017)}]{Savary:2017}%
  \BibitemOpen
  \bibfield  {author} {\bibinfo {author} {\bibfnamefont {L.}~\bibnamefont
  {Savary}}\ and\ \bibinfo {author} {\bibfnamefont {L.}~\bibnamefont
  {Balents}},\ }\href {http://stacks.iop.org/0034-4885/80/i=1/a=016502}
  {\bibfield  {journal} {\bibinfo  {journal} {Reports on Progress in Physics}\
  }\textbf {\bibinfo {volume} {80}},\ \bibinfo {pages} {016502} (\bibinfo
  {year} {2017})}\BibitemShut {NoStop}%
\bibitem [{\citenamefont {da~Silva~Neto}\ \emph {et~al.}(2013)\citenamefont
  {da~Silva~Neto}, \citenamefont {Aynajian}, \citenamefont {Baumbach},
  \citenamefont {Bauer}, \citenamefont {Mydosh}, \citenamefont {Ono},\ and\
  \citenamefont {Yazdani}}]{Neto:2013}%
  \BibitemOpen
  \bibfield  {author} {\bibinfo {author} {\bibfnamefont {E.~H.}\ \bibnamefont
  {da~Silva~Neto}}, \bibinfo {author} {\bibfnamefont {P.}~\bibnamefont
  {Aynajian}}, \bibinfo {author} {\bibfnamefont {R.~E.}\ \bibnamefont
  {Baumbach}}, \bibinfo {author} {\bibfnamefont {E.~D.}\ \bibnamefont {Bauer}},
  \bibinfo {author} {\bibfnamefont {J.}~\bibnamefont {Mydosh}}, \bibinfo
  {author} {\bibfnamefont {S.}~\bibnamefont {Ono}}, \ and\ \bibinfo {author}
  {\bibfnamefont {A.}~\bibnamefont {Yazdani}},\ }\href {\doibase
  10.1103/PhysRevB.87.161117} {\bibfield  {journal} {\bibinfo  {journal} {Phys.
  Rev. B}\ }\textbf {\bibinfo {volume} {87}},\ \bibinfo {pages} {161117}
  (\bibinfo {year} {2013})}\BibitemShut {NoStop}%
\bibitem [{\citenamefont {Rondin}\ \emph {et~al.}(2014)\citenamefont {Rondin},
  \citenamefont {Tetienne}, \citenamefont {Hingant}, \citenamefont {Roch},
  \citenamefont {Maletinsky},\ and\ \citenamefont {Jacques}}]{Rondin2014}%
  \BibitemOpen
  \bibfield  {author} {\bibinfo {author} {\bibfnamefont {L.}~\bibnamefont
  {Rondin}}, \bibinfo {author} {\bibfnamefont {J.-P.}\ \bibnamefont
  {Tetienne}}, \bibinfo {author} {\bibfnamefont {T.}~\bibnamefont {Hingant}},
  \bibinfo {author} {\bibfnamefont {J.-F.}\ \bibnamefont {Roch}}, \bibinfo
  {author} {\bibfnamefont {P.}~\bibnamefont {Maletinsky}}, \ and\ \bibinfo
  {author} {\bibfnamefont {V.}~\bibnamefont {Jacques}},\ }\href
  {http://stacks.iop.org/0034-4885/77/i=5/a=056503} {\bibfield  {journal}
  {\bibinfo  {journal} {Reports on Progress in Physics}\ }\textbf {\bibinfo
  {volume} {77}},\ \bibinfo {pages} {056503} (\bibinfo {year}
  {2014})}\BibitemShut {NoStop}%
\bibitem [{\citenamefont {Dunkin}(1998)}]{Dunkin1998}%
  \BibitemOpen
  \bibfield  {author} {\bibinfo {author} {\bibfnamefont {I.~R.}\ \bibnamefont
  {Dunkin}},\ }\href@noop {} {\emph {\bibinfo {title} {Matrix-Isolation
  Techniques}}}\ (\bibinfo  {publisher} {Oxford University Press},\ \bibinfo
  {year} {1998})\BibitemShut {NoStop}%
\bibitem [{\citenamefont {Barnes}\ \emph {et~al.}(2012)\citenamefont {Barnes},
  \citenamefont {Orville-Thomas}, \citenamefont {Gaufr\`es},\ and\
  \citenamefont {M$\ddot{\mathrm{u}}$ller}}]{matrixIsolation}%
  \BibitemOpen
  \bibfield  {author} {\bibinfo {author} {\bibfnamefont {A.}~\bibnamefont
  {Barnes}}, \bibinfo {author} {\bibfnamefont {W.}~\bibnamefont
  {Orville-Thomas}}, \bibinfo {author} {\bibfnamefont {R.}~\bibnamefont
  {Gaufr\`es}}, \ and\ \bibinfo {author} {\bibfnamefont {A.}~\bibnamefont
  {M$\ddot{\mathrm{u}}$ller}},\ }\href@noop {} {\emph {\bibinfo {title} {Matrix
  Isolation Spectroscopy}}}\ (\bibinfo  {publisher} {Springer Science \&
  Business Media},\ \bibinfo {year} {2012})\BibitemShut {NoStop}%
\bibitem [{\citenamefont {Fajardo}(1993)}]{fajardo1993}%
  \BibitemOpen
  \bibfield  {author} {\bibinfo {author} {\bibfnamefont {M.~E.}\ \bibnamefont
  {Fajardo}},\ }\href {\doibase 10.1063/1.465348} {\bibfield  {journal}
  {\bibinfo  {journal} {J. Chem. Phys}\ }\textbf {\bibinfo {volume} {99}},\
  \bibinfo {pages} {854} (\bibinfo {year} {1993})}\BibitemShut {NoStop}%
\bibitem [{\citenamefont {Ahokas}\ \emph {et~al.}(1999)\citenamefont {Ahokas},
  \citenamefont {Kiljunen}, \citenamefont {Eloranta},\ and\ \citenamefont
  {Kunttu}}]{kunttu1999}%
  \BibitemOpen
  \bibfield  {author} {\bibinfo {author} {\bibfnamefont {J.}~\bibnamefont
  {Ahokas}}, \bibinfo {author} {\bibfnamefont {T.}~\bibnamefont {Kiljunen}},
  \bibinfo {author} {\bibfnamefont {J.}~\bibnamefont {Eloranta}}, \ and\
  \bibinfo {author} {\bibfnamefont {H.}~\bibnamefont {Kunttu}},\ }\href
  {\doibase 10.1063/1.480825} {\bibfield  {journal} {\bibinfo  {journal} {J.
  Chem. Phys}\ }\textbf {\bibinfo {volume} {112}},\ \bibinfo {pages} {2420–}
  (\bibinfo {year} {1999})}\BibitemShut {NoStop}%
\bibitem [{\citenamefont {Kanagin}\ \emph {et~al.}(2013)\citenamefont
  {Kanagin}, \citenamefont {Regmi}, \citenamefont {Pathak},\ and\ \citenamefont
  {Weinstein}}]{Kanagin2013}%
  \BibitemOpen
  \bibfield  {author} {\bibinfo {author} {\bibfnamefont {A.~N.}\ \bibnamefont
  {Kanagin}}, \bibinfo {author} {\bibfnamefont {S.~K.}\ \bibnamefont {Regmi}},
  \bibinfo {author} {\bibfnamefont {P.}~\bibnamefont {Pathak}}, \ and\ \bibinfo
  {author} {\bibfnamefont {J.~D.}\ \bibnamefont {Weinstein}},\ }\href {\doibase
  10.1103/PhysRevA.88.063404} {\bibfield  {journal} {\bibinfo  {journal} {Phys.
  Rev. A}\ }\textbf {\bibinfo {volume} {88}},\ \bibinfo {pages} {063404}
  (\bibinfo {year} {2013})}\BibitemShut {NoStop}%
\bibitem [{\citenamefont {Golovizin}\ \emph {et~al.}(2015)\citenamefont
  {Golovizin}, \citenamefont {Kalganova}, \citenamefont {Sukachev},
  \citenamefont {Vishnyakova}, \citenamefont {Semerikov}, \citenamefont
  {Soshenko}, \citenamefont {Tregubov}, \citenamefont {Akimov}, \citenamefont
  {Kolachevsky}, \citenamefont {Khabarova},\ and\ \citenamefont
  {Sorokin}}]{Golovizin2015}%
  \BibitemOpen
  \bibfield  {author} {\bibinfo {author} {\bibfnamefont {A.~A.}\ \bibnamefont
  {Golovizin}}, \bibinfo {author} {\bibfnamefont {E.~S.}\ \bibnamefont
  {Kalganova}}, \bibinfo {author} {\bibfnamefont {D.~D.}\ \bibnamefont
  {Sukachev}}, \bibinfo {author} {\bibfnamefont {G.~A.}\ \bibnamefont
  {Vishnyakova}}, \bibinfo {author} {\bibfnamefont {I.~A.}\ \bibnamefont
  {Semerikov}}, \bibinfo {author} {\bibfnamefont {V.~V.}\ \bibnamefont
  {Soshenko}}, \bibinfo {author} {\bibfnamefont {D.~O.}\ \bibnamefont
  {Tregubov}}, \bibinfo {author} {\bibfnamefont {A.~V.}\ \bibnamefont
  {Akimov}}, \bibinfo {author} {\bibfnamefont {N.~N.}\ \bibnamefont
  {Kolachevsky}}, \bibinfo {author} {\bibfnamefont {K.~Y.}\ \bibnamefont
  {Khabarova}}, \ and\ \bibinfo {author} {\bibfnamefont {V.~N.}\ \bibnamefont
  {Sorokin}},\ }\href {http://stacks.iop.org/1063-7818/45/i=5/a=482} {\bibfield
   {journal} {\bibinfo  {journal} {Quantum Electronics}\ }\textbf {\bibinfo
  {volume} {45}},\ \bibinfo {pages} {482} (\bibinfo {year} {2015})}\BibitemShut
  {NoStop}%
\bibitem [{\citenamefont {Xu}\ \emph {et~al.}(2011)\citenamefont {Xu},
  \citenamefont {Hu}, \citenamefont {Singh}, \citenamefont {Bailey},
  \citenamefont {Lu}, \citenamefont {Mueller}, \citenamefont {O'Connor},\ and\
  \citenamefont {Welp}}]{Welp2011}%
  \BibitemOpen
  \bibfield  {author} {\bibinfo {author} {\bibfnamefont {C.-Y.}\ \bibnamefont
  {Xu}}, \bibinfo {author} {\bibfnamefont {S.-M.}\ \bibnamefont {Hu}}, \bibinfo
  {author} {\bibfnamefont {J.}~\bibnamefont {Singh}}, \bibinfo {author}
  {\bibfnamefont {K.}~\bibnamefont {Bailey}}, \bibinfo {author} {\bibfnamefont
  {Z.-T.}\ \bibnamefont {Lu}}, \bibinfo {author} {\bibfnamefont
  {P.}~\bibnamefont {Mueller}}, \bibinfo {author} {\bibfnamefont {T.~P.}\
  \bibnamefont {O'Connor}}, \ and\ \bibinfo {author} {\bibfnamefont
  {U.}~\bibnamefont {Welp}},\ }\href {\doibase 10.1103/PhysRevLett.107.093001}
  {\bibfield  {journal} {\bibinfo  {journal} {Phys. Rev. Lett.}\ }\textbf
  {\bibinfo {volume} {107}},\ \bibinfo {pages} {093001} (\bibinfo {year}
  {2011})}\BibitemShut {NoStop}%
\bibitem [{\citenamefont {Belyaeva}\ and\ \citenamefont
  {Predtechenskii}(1986)}]{Belyaeva1986}%
  \BibitemOpen
  \bibfield  {author} {\bibinfo {author} {\bibfnamefont {A.~A.}\ \bibnamefont
  {Belyaeva}}\ and\ \bibinfo {author} {\bibfnamefont {Y.~B.}\ \bibnamefont
  {Predtechenskii}},\ }\href@noop {} {\bibfield  {journal} {\bibinfo  {journal}
  {Opt. Spectrosc.}\ }\textbf {\bibinfo {volume} {60}},\ \bibinfo {pages} {700}
  (\bibinfo {year} {1986})}\BibitemShut {NoStop}%
\bibitem [{\citenamefont {{Martin}}\ \emph {et~al.}(1978)\citenamefont
  {{Martin}}, \citenamefont {{Zalubas}},\ and\ \citenamefont
  {{Hagan}}}]{Martin}%
  \BibitemOpen
  \bibfield  {author} {\bibinfo {author} {\bibfnamefont {W.~C.}\ \bibnamefont
  {{Martin}}}, \bibinfo {author} {\bibfnamefont {R.}~\bibnamefont {{Zalubas}}},
  \ and\ \bibinfo {author} {\bibfnamefont {L.}~\bibnamefont {{Hagan}}},\
  }\href@noop {} {\emph {\bibinfo {title} {{Atomic energy levels - The
  rare-Earth elements}}}}\ (\bibinfo  {publisher} {NSRDS-NBS, Washington:
  National Bureau of Standards, U.S.~Department of Commerce},\ \bibinfo {year}
  {1978})\BibitemShut {NoStop}%
\bibitem [{\citenamefont {Aleksandrov}\ \emph {et~al.}(1984)\citenamefont
  {Aleksandrov}, \citenamefont {Vedenin},\ and\ \citenamefont
  {Kulyasov}}]{Aleksandrov1984}%
  \BibitemOpen
  \bibfield  {author} {\bibinfo {author} {\bibfnamefont {E.~B.}\ \bibnamefont
  {Aleksandrov}}, \bibinfo {author} {\bibfnamefont {V.~D.}\ \bibnamefont
  {Vedenin}}, \ and\ \bibinfo {author} {\bibfnamefont {V.~N.}\ \bibnamefont
  {Kulyasov}},\ }\href@noop {} {\bibfield  {journal} {\bibinfo  {journal} {Opt.
  Spectrosc.}\ }\textbf {\bibinfo {volume} {56}},\ \bibinfo {pages} {365}
  (\bibinfo {year} {1984})}\BibitemShut {NoStop}%
\bibitem [{\citenamefont {Connolly}\ \emph {et~al.}(2010)\citenamefont
  {Connolly}, \citenamefont {Au}, \citenamefont {Doret}, \citenamefont
  {Ketterle},\ and\ \citenamefont {Doyle}}]{Connolly2010}%
  \BibitemOpen
  \bibfield  {author} {\bibinfo {author} {\bibfnamefont {C.~B.}\ \bibnamefont
  {Connolly}}, \bibinfo {author} {\bibfnamefont {Y.~S.}\ \bibnamefont {Au}},
  \bibinfo {author} {\bibfnamefont {S.~C.}\ \bibnamefont {Doret}}, \bibinfo
  {author} {\bibfnamefont {W.}~\bibnamefont {Ketterle}}, \ and\ \bibinfo
  {author} {\bibfnamefont {J.~M.}\ \bibnamefont {Doyle}},\ }\href {\doibase
  10.1103/PhysRevA.81.010702} {\bibfield  {journal} {\bibinfo  {journal} {Phys.
  Rev. A}\ }\textbf {\bibinfo {volume} {81}},\ \bibinfo {pages} {010702}
  (\bibinfo {year} {2010})}\BibitemShut {NoStop}%
\bibitem [{\citenamefont {Sukachev}\ \emph {et~al.}(2010)\citenamefont
  {Sukachev}, \citenamefont {Sokolov}, \citenamefont {Chebakov}, \citenamefont
  {Akimov}, \citenamefont {Kanorsky}, \citenamefont {Kolachevsky},\ and\
  \citenamefont {Sorokin}}]{Sukachev2010}%
  \BibitemOpen
  \bibfield  {author} {\bibinfo {author} {\bibfnamefont {D.}~\bibnamefont
  {Sukachev}}, \bibinfo {author} {\bibfnamefont {A.}~\bibnamefont {Sokolov}},
  \bibinfo {author} {\bibfnamefont {K.}~\bibnamefont {Chebakov}}, \bibinfo
  {author} {\bibfnamefont {A.}~\bibnamefont {Akimov}}, \bibinfo {author}
  {\bibfnamefont {S.}~\bibnamefont {Kanorsky}}, \bibinfo {author}
  {\bibfnamefont {N.}~\bibnamefont {Kolachevsky}}, \ and\ \bibinfo {author}
  {\bibfnamefont {V.}~\bibnamefont {Sorokin}},\ }\href {\doibase
  10.1103/PhysRevA.82.011405} {\bibfield  {journal} {\bibinfo  {journal} {Phys.
  Rev. A}\ }\textbf {\bibinfo {volume} {82}},\ \bibinfo {pages} {011405}
  (\bibinfo {year} {2010})}\BibitemShut {NoStop}%
\bibitem [{\citenamefont {Sukachev}\ \emph {et~al.}(2014)\citenamefont
  {Sukachev}, \citenamefont {Kalganova}, \citenamefont {Sokolov}, \citenamefont
  {Fedorov}, \citenamefont {Vishnyakova}, \citenamefont {Akimov}, \citenamefont
  {Kolachevsky},\ and\ \citenamefont {Sorokin}}]{Sukachev2014}%
  \BibitemOpen
  \bibfield  {author} {\bibinfo {author} {\bibfnamefont {D.~D.}\ \bibnamefont
  {Sukachev}}, \bibinfo {author} {\bibfnamefont {E.~S.}\ \bibnamefont
  {Kalganova}}, \bibinfo {author} {\bibfnamefont {A.~V.}\ \bibnamefont
  {Sokolov}}, \bibinfo {author} {\bibfnamefont {S.~A.}\ \bibnamefont
  {Fedorov}}, \bibinfo {author} {\bibfnamefont {G.~A.}\ \bibnamefont
  {Vishnyakova}}, \bibinfo {author} {\bibfnamefont {A.~V.}\ \bibnamefont
  {Akimov}}, \bibinfo {author} {\bibfnamefont {N.~N.}\ \bibnamefont
  {Kolachevsky}}, \ and\ \bibinfo {author} {\bibfnamefont {V.~N.}\ \bibnamefont
  {Sorokin}},\ }\href {http://stacks.iop.org/1063-7818/44/i=6/a=515} {\bibfield
   {journal} {\bibinfo  {journal} {Quantum Electronics}\ }\textbf {\bibinfo
  {volume} {44}},\ \bibinfo {pages} {515} (\bibinfo {year} {2014})}\BibitemShut
  {NoStop}%
\bibitem [{\citenamefont {Kalganova}\ \emph {et~al.}(2017)\citenamefont
  {Kalganova}, \citenamefont {Prudnikov}, \citenamefont {Vishnyakova},
  \citenamefont {Golovizin}, \citenamefont {Tregubov}, \citenamefont
  {Sukachev}, \citenamefont {Khabarova}, \citenamefont {Sorokin},\ and\
  \citenamefont {Kolachevsky}}]{Kalganova2017}%
  \BibitemOpen
  \bibfield  {author} {\bibinfo {author} {\bibfnamefont {E.}~\bibnamefont
  {Kalganova}}, \bibinfo {author} {\bibfnamefont {O.}~\bibnamefont
  {Prudnikov}}, \bibinfo {author} {\bibfnamefont {G.}~\bibnamefont
  {Vishnyakova}}, \bibinfo {author} {\bibfnamefont {A.}~\bibnamefont
  {Golovizin}}, \bibinfo {author} {\bibfnamefont {D.}~\bibnamefont {Tregubov}},
  \bibinfo {author} {\bibfnamefont {D.}~\bibnamefont {Sukachev}}, \bibinfo
  {author} {\bibfnamefont {K.}~\bibnamefont {Khabarova}}, \bibinfo {author}
  {\bibfnamefont {V.}~\bibnamefont {Sorokin}}, \ and\ \bibinfo {author}
  {\bibfnamefont {N.}~\bibnamefont {Kolachevsky}},\ }\href {\doibase
  10.1103/PhysRevA.96.033418} {\bibfield  {journal} {\bibinfo  {journal} {Phys.
  Rev. A}\ }\textbf {\bibinfo {volume} {96}},\ \bibinfo {pages} {033418}
  (\bibinfo {year} {2017})}\BibitemShut {NoStop}%
\bibitem [{\citenamefont {Sukachev}\ \emph {et~al.}(2016)\citenamefont
  {Sukachev}, \citenamefont {Fedorov}, \citenamefont {Tolstikhina},
  \citenamefont {Tregubov}, \citenamefont {Kalganova}, \citenamefont
  {Vishnyakova}, \citenamefont {Golovizin}, \citenamefont {Kolachevsky},
  \citenamefont {Khabarova},\ and\ \citenamefont {Sorokin}}]{Sukachev2016}%
  \BibitemOpen
  \bibfield  {author} {\bibinfo {author} {\bibfnamefont {D.}~\bibnamefont
  {Sukachev}}, \bibinfo {author} {\bibfnamefont {S.}~\bibnamefont {Fedorov}},
  \bibinfo {author} {\bibfnamefont {I.}~\bibnamefont {Tolstikhina}}, \bibinfo
  {author} {\bibfnamefont {D.}~\bibnamefont {Tregubov}}, \bibinfo {author}
  {\bibfnamefont {E.}~\bibnamefont {Kalganova}}, \bibinfo {author}
  {\bibfnamefont {G.}~\bibnamefont {Vishnyakova}}, \bibinfo {author}
  {\bibfnamefont {A.}~\bibnamefont {Golovizin}}, \bibinfo {author}
  {\bibfnamefont {N.}~\bibnamefont {Kolachevsky}}, \bibinfo {author}
  {\bibfnamefont {K.}~\bibnamefont {Khabarova}}, \ and\ \bibinfo {author}
  {\bibfnamefont {V.}~\bibnamefont {Sorokin}},\ }\href {\doibase
  10.1103/PhysRevA.94.022512} {\bibfield  {journal} {\bibinfo  {journal} {Phys.
  Rev. A}\ }\textbf {\bibinfo {volume} {94}},\ \bibinfo {pages} {022512}
  (\bibinfo {year} {2016})}\BibitemShut {NoStop}%
\bibitem [{\citenamefont {Vishnyakova}\ \emph {et~al.}(2016)\citenamefont
  {Vishnyakova}, \citenamefont {Golovizin}, \citenamefont {Kalganova},
  \citenamefont {Sorokin}, \citenamefont {Sukachev}, \citenamefont {Tregubov},
  \citenamefont {Khabarova},\ and\ \citenamefont
  {Kolachevsky}}]{Vishnyakova2016}%
  \BibitemOpen
  \bibfield  {author} {\bibinfo {author} {\bibfnamefont {G.~A.}\ \bibnamefont
  {Vishnyakova}}, \bibinfo {author} {\bibfnamefont {A.~A.}\ \bibnamefont
  {Golovizin}}, \bibinfo {author} {\bibfnamefont {E.~S.}\ \bibnamefont
  {Kalganova}}, \bibinfo {author} {\bibfnamefont {V.~N.}\ \bibnamefont
  {Sorokin}}, \bibinfo {author} {\bibfnamefont {D.~D.}\ \bibnamefont
  {Sukachev}}, \bibinfo {author} {\bibfnamefont {D.~O.}\ \bibnamefont
  {Tregubov}}, \bibinfo {author} {\bibfnamefont {K.~Y.}\ \bibnamefont
  {Khabarova}}, \ and\ \bibinfo {author} {\bibfnamefont {N.~N.}\ \bibnamefont
  {Kolachevsky}},\ }\href {http://stacks.iop.org/1063-7869/59/i=2/a=168}
  {\bibfield  {journal} {\bibinfo  {journal} {Physics-Uspekhi}\ }\textbf
  {\bibinfo {volume} {59}},\ \bibinfo {pages} {168} (\bibinfo {year}
  {2016})}\BibitemShut {NoStop}%
\bibitem [{\citenamefont {Kramida}\ \emph {et~al.}(2018)\citenamefont
  {Kramida}, \citenamefont {{Yu.~Ralchenko}}, \citenamefont {Reader},\ and\
  \citenamefont {{and NIST ASD Team}}}]{NIST_ASD}%
  \BibitemOpen
  \bibfield  {author} {\bibinfo {author} {\bibfnamefont {A.}~\bibnamefont
  {Kramida}}, \bibinfo {author} {\bibnamefont {{Yu.~Ralchenko}}}, \bibinfo
  {author} {\bibfnamefont {J.}~\bibnamefont {Reader}}, \ and\ \bibinfo {author}
  {\bibnamefont {{and NIST ASD Team}}},\ }\href@noop {} {}\bibinfo
  {howpublished} {{NIST Atomic Spectra Database (ver. 5.5.6), [Online].
  Available: {\tt{https://physics.nist.gov/asd}} [2018, November 6]. National
  Institute of Standards and Technology, Gaithersburg, MD.}} (\bibinfo {year}
  {2018})\BibitemShut {NoStop}%
\bibitem [{\citenamefont {Miller}\ and\ \citenamefont
  {Sharp}(1970)}]{Miller:1970hrba}%
  \BibitemOpen
  \bibfield  {author} {\bibinfo {author} {\bibfnamefont {J.~E.}\ \bibnamefont
  {Miller}}\ and\ \bibinfo {author} {\bibfnamefont {E.~J.}\ \bibnamefont
  {Sharp}},\ }\href {https://doi.org/10.1063/1.1658520} {\bibfield  {journal}
  {\bibinfo  {journal} {Journal of Applied Physics}\ }\textbf {\bibinfo
  {volume} {41}},\ \bibinfo {pages} {4718} (\bibinfo {year}
  {1970})}\BibitemShut {NoStop}%
\bibitem [{\citenamefont {Bespalov}\ \emph {et~al.}(1997)\citenamefont
  {Bespalov}, \citenamefont {Kazakov}, \citenamefont {Leushin},\ and\
  \citenamefont {Safiullin}}]{Bespalov:1997kzba}%
  \BibitemOpen
  \bibfield  {author} {\bibinfo {author} {\bibfnamefont {V.~F.}\ \bibnamefont
  {Bespalov}}, \bibinfo {author} {\bibfnamefont {B.~N.}\ \bibnamefont
  {Kazakov}}, \bibinfo {author} {\bibfnamefont {A.~M.}\ \bibnamefont
  {Leushin}}, \ and\ \bibinfo {author} {\bibfnamefont {G.~M.}\ \bibnamefont
  {Safiullin}},\ }\href@noop {} {\bibfield  {journal} {\bibinfo  {journal}
  {Physics of the Solid State}\ }\textbf {\bibinfo {volume} {39}},\ \bibinfo
  {pages} {925} (\bibinfo {year} {1997})}\BibitemShut {NoStop}%
\bibitem [{\citenamefont {Sugiyama}\ \emph {et~al.}(2006)\citenamefont
  {Sugiyama}, \citenamefont {Katsurayama}, \citenamefont {Anzai},\ and\
  \citenamefont {Tsuboi}}]{Sugiyama:2006chba}%
  \BibitemOpen
  \bibfield  {author} {\bibinfo {author} {\bibfnamefont {A.}~\bibnamefont
  {Sugiyama}}, \bibinfo {author} {\bibfnamefont {M.}~\bibnamefont
  {Katsurayama}}, \bibinfo {author} {\bibfnamefont {Y.}~\bibnamefont {Anzai}},
  \ and\ \bibinfo {author} {\bibfnamefont {T.}~\bibnamefont {Tsuboi}},\
  }\href@noop {} {\bibfield  {journal} {\bibinfo  {journal} {Proceedings of
  Rare Earths'04 in Nara, Japan}\ }\textbf {\bibinfo {volume} {408-412}},\
  \bibinfo {pages} {780} (\bibinfo {year} {2006})}\BibitemShut {NoStop}%
\bibitem [{\citenamefont {Gerhardt}\ \emph {et~al.}(2012)\citenamefont
  {Gerhardt}, \citenamefont {Sin},\ and\ \citenamefont {Momose}}]{Ilja:2012}%
  \BibitemOpen
  \bibfield  {author} {\bibinfo {author} {\bibfnamefont {I.}~\bibnamefont
  {Gerhardt}}, \bibinfo {author} {\bibfnamefont {K.}~\bibnamefont {Sin}}, \
  and\ \bibinfo {author} {\bibfnamefont {T.}~\bibnamefont {Momose}},\ }\href
  {\doibase 10.1063/1.4730032} {\bibfield  {journal} {\bibinfo  {journal} {The
  Journal of Chemical Physics}\ }\textbf {\bibinfo {volume} {137}},\ \bibinfo
  {pages} {014507} (\bibinfo {year} {2012})},\ \Eprint
  {http://arxiv.org/abs/https://doi.org/10.1063/1.4730032}
  {https://doi.org/10.1063/1.4730032} \BibitemShut {NoStop}%
\bibitem [{\citenamefont {Upadhyay}\ \emph {et~al.}(2016)\citenamefont
  {Upadhyay}, \citenamefont {Kanagin}, \citenamefont {Hartzell}, \citenamefont
  {Christy}, \citenamefont {Arnott}, \citenamefont {Momose}, \citenamefont
  {Patterson},\ and\ \citenamefont {Weinstein}}]{upadhyay2016}%
  \BibitemOpen
  \bibfield  {author} {\bibinfo {author} {\bibfnamefont {S.}~\bibnamefont
  {Upadhyay}}, \bibinfo {author} {\bibfnamefont {A.~N.}\ \bibnamefont
  {Kanagin}}, \bibinfo {author} {\bibfnamefont {C.}~\bibnamefont {Hartzell}},
  \bibinfo {author} {\bibfnamefont {T.}~\bibnamefont {Christy}}, \bibinfo
  {author} {\bibfnamefont {W.~P.}\ \bibnamefont {Arnott}}, \bibinfo {author}
  {\bibfnamefont {T.}~\bibnamefont {Momose}}, \bibinfo {author} {\bibfnamefont
  {D.}~\bibnamefont {Patterson}}, \ and\ \bibinfo {author} {\bibfnamefont
  {J.~D.}\ \bibnamefont {Weinstein}},\ }\href {\doibase
  10.1103/PhysRevLett.117.175301} {\bibfield  {journal} {\bibinfo  {journal}
  {Phys. Rev. Lett.}\ }\textbf {\bibinfo {volume} {117}},\ \bibinfo {pages}
  {175301} (\bibinfo {year} {2016})}\BibitemShut {NoStop}%
\bibitem [{\citenamefont {Epstein}\ \emph {et~al.}(1995)\citenamefont
  {Epstein}, \citenamefont {Buchwald}, \citenamefont {Edwards}, \citenamefont
  {Gosnell},\ and\ \citenamefont {Mungan}}]{Epstein1995}%
  \BibitemOpen
  \bibfield  {author} {\bibinfo {author} {\bibfnamefont {R.~I.}\ \bibnamefont
  {Epstein}}, \bibinfo {author} {\bibfnamefont {M.~I.}\ \bibnamefont
  {Buchwald}}, \bibinfo {author} {\bibfnamefont {B.~C.}\ \bibnamefont
  {Edwards}}, \bibinfo {author} {\bibfnamefont {T.~R.}\ \bibnamefont
  {Gosnell}}, \ and\ \bibinfo {author} {\bibfnamefont {C.~E.}\ \bibnamefont
  {Mungan}},\ }\href {https://doi.org/10.1038/377500a0} {\bibfield  {journal}
  {\bibinfo  {journal} {Nature}\ }\textbf {\bibinfo {volume} {377}},\ \bibinfo
  {pages} {500 EP } (\bibinfo {year} {1995})}\BibitemShut {NoStop}%
\bibitem [{\citenamefont {Seletskiy}\ \emph {et~al.}(2016)\citenamefont
  {Seletskiy}, \citenamefont {Epstein},\ and\ \citenamefont
  {Sheik-Bahae}}]{Sheik-Bahae2016}%
  \BibitemOpen
  \bibfield  {author} {\bibinfo {author} {\bibfnamefont {D.~V.}\ \bibnamefont
  {Seletskiy}}, \bibinfo {author} {\bibfnamefont {R.}~\bibnamefont {Epstein}},
  \ and\ \bibinfo {author} {\bibfnamefont {M.}~\bibnamefont {Sheik-Bahae}},\
  }\href {http://stacks.iop.org/0034-4885/79/i=9/a=096401} {\bibfield
  {journal} {\bibinfo  {journal} {Reports on Progress in Physics}\ }\textbf
  {\bibinfo {volume} {79}},\ \bibinfo {pages} {096401} (\bibinfo {year}
  {2016})}\BibitemShut {NoStop}%
\end{thebibliography}
\end{document}